\documentclass[aps,prd,twocolumn,superscriptaddress,groupedaddress]{revtex4}  
\usepackage{graphicx}  % needed for figures
\usepackage{dcolumn}   % needed for some tables
\usepackage{bm}        % for math
\usepackage{amssymb}   % for math

\newcommand{\be}{\begin{equation}}
\newcommand{\ee}{\end{equation}}
\newcommand{\bea}{\begin{eqnarray}}
\newcommand{\eea}{\end{eqnarray}}
\newcommand{\bg}{\begin{gather}}
\newcommand{\eg}{\end{gather}}
\newcommand{\bseq}{\begin{subequations}}
\newcommand{\eseq}{\end{subequations}}

\newcommand{\seq}{\begin{subequations}}
\newcommand{\sen}{\end{subequations}}
\newcommand{\eq}{\begin{eqnarray}}
\newcommand{\en}{\end{eqnarray}}

\def\shiftdown#1{#1\llap{\lower.04ex\hbox{#1}}}

\begin{document}

\title{Implication of the hidden sub-GeV bosons for the  $(g-2)_\mu$,  
$^8$Be-$^4$He anomaly, \\ proton charge radius, EDM of fermions 
and dark axion portal}
\author{D.~V.~Kirpichnikov\footnote{{\bf e-mail}: kirpich@ms2.inr.ac.ru}}
\affiliation{Institute for Nuclear Research of the Russian Academy 
of Sciences, 117312 Moscow, Russia} 
\author{Valery~E.~Lyubovitskij\footnote{{\bf e-mail}: 
valery.lyubovitsky@cern.ch}} 
\affiliation{Institut f\"ur Theoretische Physik,
Universit\"at T\"ubingen,
Kepler Center for Astro and Particle Physics,
Auf der Morgenstelle 14, D-72076 T\"ubingen, Germany}
\affiliation{Departamento de F\'\i sica y Centro Cient\'\i fico
Tecnol\'ogico de Valpara\'\i so-CCTVal, Universidad T\'ecnica
Federico Santa Mar\'\i a, Casilla 110-V, Valpara\'\i so, Chile}
\affiliation{Department of Physics, Tomsk State University,
634050 Tomsk, Russia}
\affiliation{Tomsk Polytechnic University, 634050 Tomsk, Russia}
\author{Alexey S.~Zhevlakov\footnote{{\bf e-mail}:
zhevlakov1@gmail.com}}
\affiliation{Department of Physics, Tomsk State University,
634050 Tomsk, Russia}
\affiliation{Matrosov Institute for System Dynamics and 
Control Theory SB RAS Lermontov str., 134, 664033, Irkutsk, Russia}
                                                        
\date{\today}

\begin{abstract}
 We discuss new physics phenomenology of hidden scalar ($S$), pseudoscalar ($P$),  vector ($V$) and axial-vector ($A$) 
particles coupled to nucleons and leptons, which 
could give contributions to proton charge radius, $(g-2)_\mu$, 
$^8$Be-$^4$He anomaly and electric dipole moment (EDM) 
of  Standard Model (SM)  particles. In particular, we  estimate sensitivity of 
NA64$\mu$ experiment to observe muon missing energy events involving hidden scalar
and vector particles. That analysis is based on {\tt GEANT4} Monte Carlo simulation 
of the signal process of muon scattering off target nuclei $\mu N \to \mu N S(V)$ followed by invisible boson decay into Dark Matter 
(DM) particles, $S(V)\to \chi \chi$. The existence of light sub-GeV bosons could possibly 
explain the muon $(g-2)$ anomaly observed. We also summarize existing bounds on ATOMKI 
$X17(J^P=0^-, 1^\pm)$  
boson coupling with neutron, proton and electron. We implement these constraints to estimate the
contribution of $P$, $V$ and $A$ particles to proton charge radius
via direct 1-loop calculation of  Sachs form factors. The analysis reveals the corresponding
contribution is negligible. We also 
calculate bounds on dark axion portal couplings of dimension-five operators, which contribute to
the EDMs of leptons and neutron.
\end{abstract}

\maketitle

\section{Introduction
\label{IntroSec}}
The measurement of anomalous magnetic moment of muon provides the 
potential signal of new physics. Indeed, the value of $(g-2)_\mu$ 
measured by BNL~\cite{Bennett:2006fi} differs from the prediction 
of Standard Model (SM) at the level of $3.7$ standard 
deviations~\cite{Blum:2019ugy,Aoyama:2020ynm}, 
$\Delta a_\mu =a_{\mu}^{exp}- a_{\mu}^{th}=(279\pm 
76)\times 10^{-11}$. The existence of light and weakly coupled hidden 
bosons~\cite{fayet1,fayet2,fayet3,fayet4}
could be a possible beyond SM explanations of that 
discrepancy~\cite{Alekhin:2015byh,Alexander:2016aln}. In particular,
{\tt BELLEII} experiment~\cite{Adachi:2019otg} has been already 
put constraints on hidden
vector boson $Z'$ coupled with muons, which can contribute to $(g-2)_\mu$ 
anomaly. In addition, $M^3$ compact muon missing momentum 
experiment~\cite{Kahn:2018cqs}
has been proposed recently at Fermilab to examine $(g-2)_\mu$ puzzle.
Moreover, muon (electron) fixed target 
NA64$\mu$ (NA64$e$) experiment at CERN SPS~\cite{NA64muProposal} plans 
to collect  data after CERN Long Shutdown (LS2) in~2021 to test sub-GeV 
boson contribution  into muon and electron $(g-2)$. In particular, 
the NA64 experiment at the CERN SPS 
combines the active target and missing energy techniques  
to   search for  rare events.

The  processes accompanied by the emission and decay of 
hypothetical hidden boson~\cite{Krasznahorkay:2015iga,Krasznahorkay:2019lyl} 
provide an  additional evidence towards the weakly coupled particle 
interactions 
beyond SM~\cite{Feng:2016jff,Feng:2016ysn,Ellwanger:2016wfe,
Kozaczuk:2016nma,Banerjee:2019hmi}. Namely, ATOMKI Collaboration has been 
reported recently the $\sim 6.8 \sigma$ and $\sim 7.2 \sigma$ anomalies of 
$e^+e^-$ pair excess from electro-magnetically transition in 
$^8$Be~\cite{Krasznahorkay:2015iga} and $^4$He~\cite{Krasznahorkay:2019lyl}, 
respectively. The relevant $^8$Be data have been explained as creation and 
decay of $X17$ boson particle with mass $m_X= 16.70\pm 0.35\pm 0.50$~MeV. 
Furthermore, most favored candidates, that could play the role  of the 
$X17$  boson~\cite{Feng:2016ysn,Ellwanger:2016wfe,Kozaczuk:2016nma} have 
spin-parity $J^P = 1^+$, $J^P=0^-$, and $J^P=1^-$.  
In particular, in order to explain $^8$Be anomaly, authors 
of Ref.~\cite{Feng:2016ysn} provided an analysis for excited $^8$Be states 
and presented anomaly-free extension of SM that contains gauge boson with 
experimentally favored couplings~\cite{Banerjee:2019hmi,Banerjee:2018vgk,%
Raggi:2015noa,Batley:2015lha,Blumlein:2013cua,Blumlein:2011mv,%
Davoudiasl:2014kua,Babusci:2012cr,Adlarson:2013eza,Agakishiev:2013fwl}. 
In addition, in Ref.~\cite{Ellwanger:2016wfe} light 
pseudo-scalar state from Higgs extended sector was suggested to describe 
relevant $e^+e^-$ excess in $^8$Be transition with coupling that satisfies    
existing constraints~\cite{Andreas:2010ms,Adler:2004hp,Artamonov:2009sz}. 
Moreover, authors of Ref.~\cite{Kozaczuk:2016nma} 
investigated the production of vector boson with primarily axial couplings 
to quarks that is consistent with experimental 
data~\cite{Ginges:2003qt,Wood:1997zq,Bouchiat:2004sp}, such that new axial 
field has a mass $m_X \simeq 16.7$ MeV (see e.g., 
Refs.~\cite{Kahn:2016vjr,DelleRose:2017xil,DelleRose:2018pgm} for recent 
review) and describes comprehensively nuclear properties of 
the $^8$Be$(1^+)\to ^8$Be$(0^+)$ anomalous transition.

However, in~\cite{Zhang:2017zap} authors provide  dedicated analysis of  
$e^+e^-$ pair emission anisotropy in nuclear  transitions of $^8$Be,  
which has a possible relevance to that anomaly. Another  analysis 
of $^8$Be anomaly not involving beyond Standard Model explanation was 
carried out recently in Ref.~\cite{Tursunov:2020wfy}. In particular, 
 author provides a hint that 
$17$~MeV excess in the experiment with $^8$Be~\cite{Krasznahorkay:2015iga} 
and $^8$He~\cite{Krasznahorkay:2019lyl} can be associated
 with the quantum phase  transition in the $\alpha$-like 
nuclei of $^8$Be, $^4$He, $^{12}$C, and $^{16}$O.
% Moreover, it
%would be instructive to demonstrate an explicit calculation of angle 
%distribution of $e^+e^-$ excess for that scenario.    

It is worth mentioning  
that electron fixed target NA64e facility at the CERN SPS~\cite{Banerjee:2019hmi,Banerjee:2018vgk} 
has excellent opportunity of probing 
$^8$Be anomaly due to its dedicated searching sensitivity for short-lived 
hidden particles, $ \tau_X \lesssim 10^{-12}$ s. In particular, we expect that 
NA64e active target facility will be able to probe hidden pseudoscalar
$X17$ boson after CERN LS$2$ in~$2021$.  
 
Precise determination of the proton charge radius $r_p^E$, 
one of the fundamental 
quantities of hadron physics, remains unsolved problem for many years. 
There are three methods of measurement of the proton charge radius 
from study: (1) cross section of elastic lepton-proton scattering, 
(2) Lamb shift in atomic hydrogen, and 
(3) Lamb shift in muonic hydrogen. 

The most recent and precise result for the $r_p^E$ extracted from 
the elastic electron scattering off proton was  
obtained by the A1 Collaboration at MAMI~\cite{Bernauer:2010wm}: 
$r_p^E = 0.879 \pm 0.005 \pm 0.006$ fm. It is in a good agreement 
with the 2014 CODATA recommended value 
$r_p^E = 0.8751 \pm 0.0061$ fm~\cite{Mohr:2015ccw}. 
However, these results are in a sizable disagreement (by 5.6 
standard deviations) with most accurate result for 
the $r_p^E = 0.84087 \pm 0.00026 \pm 0.00029$ fm obtained from 
Lamb shift in $\mu p$ atom by the CREMA Collaboration 
at PSI~\cite{Antognini:1900ns,Pohl:2010zza}. 
In 2019 the proton radius was deduce from measurement 
of the electronic hydrogen Lamb shift: 
$r_p^E = 0.833 \pm 0.010$ fm~\cite{Bezginov:2019mdi}, which 
led to a conclusion that  
the electron- and muon-based measurements of the $r_p^E$ 
finally agrees with each other. 
Recently, the PRad Collaboration at JLab~\cite{Xiong:2019umf} 
reported on improved measurement of the proton charge radius 
from an electron-proton scattering experiment: 
$r_p^E = 0.831 \pm 0.007({\rm stat}) \pm 0.012({\rm syst})$ fm. 
As stressed in Ref.~\cite{Xiong:2019umf}, this prediction is 
smaller than the most recent high-accuracy predictions based 
on $ep$ elastic scattering and very close to the results of the 
precise muonic hydrogen experiments~\cite{Pohl:2010zza,Antognini:1900ns}. 
Also it was noticed in~\cite{Xiong:2019umf} that their prediction is 
2.7 standard deviations smaller than the average of all $ep$ experimental 
results~\cite{Mohr:2015ccw}. We note that an independent and a highly-precise
measurement proposed by the COMPASS++/AMBER at the M2 beam line of 
the CERN SPS~\cite{Denisov:2018unj} has very strong physical motivation 
as independent and complimentary experiment to recent observation done 
by the PRad Collaboration~\cite{Xiong:2019umf}. On the other hand, 
the use of the muon beam in the planned COMPASS++/AMBER 
experiment~\cite{Denisov:2018unj} 
gives a unique opportunity to test electron-muon universality 
and to reduce systematic uncertainties and radiative 
corrections. For discussion of  future experiments and overview on 
proton radius see, e.g., Refs.~\cite{Carlson:2015jba,Hammer:2019uab,
Lorenzon:2020qsh}.

One should stress that from theoretical point new particles 
with different spin-parity assignments could contribute to resolving 
of puzzles in particle phenomenology and to more precise 
determination of their properties. E.g., one can imagine existence 
of new particles with different spin-parity assignments, 
e.g., scalar ($J^P =0^+$), pseudoscalar ($J^P =0^-$), 
vector ($J^P =1^+$), and axial ($J^P =1^-$) particles. 
Also one can analyze a possible contribution 
of these states to the $(g-2)_\mu$ anomaly. Note that effects of 
scalar, pseudoscalar, and vector particles on the Lamb shift 
in lepton-hydrogen 
and  $(g-2)_\mu$ anomaly have been already discussed and estimated 
in literature~\cite{Jackiw:1972jz}-\cite{Liu:2016qwd}. 
We noticed that one can also estimate the relative contribution of new 
particles ($S$, $P$, $V$, and $A$) 
to the proton charge radius via direct 1-loop calculation of Sachs form factors. 
From our preliminary analysis it follows that 
contribution of these particles to the charge radius of proton 
is negligible. 

 However, it is instructive to collect existing bounds on $X17$ boson coupling with SM 
fermions  and calculate  contribution of $X17$ to EDMs of leptons and neutron. 
The relevant coupling terms originate from dimension-five operators 
(see e.g., Eq.~(\ref{AxionPortalCouplings}) below). These interactions are motivated by 
dark-axion portal scenarios, 
involving couplings of  photon, dark photon, and axion-like 
particle (for details, see e.~g.,~ 
Refs.~\cite{Kaneta:2016wvf,Kaneta:2017wfh,Hochberg:2018rjs,
deNiverville:2019xsx,deNiverville:2018hrc,Daido:2019tbm}). In 
addition, several well motivated scenarios of new physics involving the light hidden sector  and EDMs are discussed in Ref.~\cite{Dall:2015bba,Okada:2019sbb,Yamanaka:2017mef,Yanase:2018qqq}.

Our paper is structured as follows. 
In Sec.~\ref{EffLagrSect} we consider effective couplings of sub-GeV 
bosons with SM fermions. 
In Sec.~\ref{NA64muSect} we estimate sensitivity of NA64$\mu$ muon active 
target experiment to probe sub-GeV Vector and Scalar mediator of DM by using 
comprehensive {\tt GEANT4} MC simulation. These bosons can possibly
explain $(g-2)_\mu$ anomaly.
In Sec.~\ref{8BeSect} we summarize existing constraints on $^8$Be anomaly 
for hidden $X17(J^P=0^-,1^\pm)$ bosons.
In Sec.~\ref{EDMSect} we estimate contribution of $X17(J^P=0^-,1^\pm)$ bosons 
to proton charge radius directly from Sachs form factors. 
We conclude, that current information on new 
particles suggests that their contribution to the charge radius of proton 
is negligible. 
In Sec.~\ref{EDMSect} we also set constraints on dimension-five operator 
couplings of light bosons which can contribute to EDM of SM fermions. 
That analysis is motivated by dark axion portal study.  
Finally, in Sec.~\ref{summary}, we summarize our results.

\section{Effective Lagrangian
\label{EffLagrSect}}
 We consider entirely phenomenological couplings of light bosons 
to SM particles, which are based on an effective theory approach. Namely,
New Physics (NP) Lagrangian involving coupling of nucleons
and leptons with scalar $S$, pseudoscalar $P$, vector $V$,
and axial $A$ bosons, which could contribute to the proton radius,
muon magnetic moment, and electric dipole moments of electron (muon)
and neutron   can be written as follows
\eq
{\cal L}_{\rm NP} =
\sum\limits_{H} {\cal L}_H \,+\,
\sum\limits_{H_1H_2} {\cal L}_{\gamma H_1H_2} \,,
\label{LagrTot}
\en
where $H=S,P,V,A$ and $H_1H_2=SV,PV,PA$. Here
${\cal L}_H = H J_H$, 
where $J_H$ is the fermionic currents including effects
of $P$-parity violation. They are
composed of nucleons and fermions as
\eq\label{NuclAndLept}  
J_H &=& \sum\limits_{N=p, n}
\, \bar N\, (g_H^N \, \Gamma_H \,+\,
f_H^N \, \tilde\Gamma_H) \, N \nonumber\\
&+&\sum\limits_{\ell=e, \mu, \tau} \,
\, \bar\ell \, (g_H^\ell \, \Gamma_H \,+\,
f_H^\ell \, \tilde\Gamma_H) \, \ell \,,
\en
where $\Gamma_S = \tilde\Gamma_P = I$,
      $\Gamma_P = \tilde\Gamma_S = i\gamma^5$,
      $\Gamma_V = \tilde\Gamma_A = \gamma^\mu$, and
      $\Gamma_A = \tilde\Gamma_V = \gamma^\mu\gamma^5$
are the Dirac spin matrices.  
Second term in Lagrangian (\ref{LagrTot})
describes the coupling of new particles with photon 
(here we list only the terms which contribute to 
the electric dipole moment):
\eq
{\cal L}_{\gamma SV} &=& \frac{e}{4M_p} \, 
g_{\gamma SV} \, F^{\mu\nu}\, V_{\mu\nu} \, S  \,,
\nonumber\\
{\cal L}_{\gamma PA} &=& \frac{e}{4M_p} \,
g_{\gamma PA} \, F^{\mu\nu} \, A_{\mu\nu} \, P  \,,
\nonumber\\
{\cal L}_{\gamma PV} &=& \frac{e}{4M_p} \,
f_{\gamma PV} \, F^{\mu\nu} \, V_{\mu\nu} \, P  \,. 
\label{AxionPortalCouplings}
\en
$g_H^{N(\ell)}$, $g_{\gamma H_1H_2}$
and $f_H^{N(\ell)}$, $f_{\gamma H_1H_2}$
are the sets of $P$-parity even and $P$-parity
odd couplings, respectively.
In Appendix we list the expressions for the contributions
of new particles to the muon magnetic moment and proton charge radius
including both $P$-even and $P$-odd couplings, while in numerical
analysis, for simplicity we will neglect by the $P$-odd couplings.
Later, we derive the constraints of combinations of $P$-even and
$P$-odd couplings of new particles using data on electric dipole
moments of leptons and neutron. However,  
we note that constraints on (\ref{AxionPortalCouplings}) 
couplings can be motivated by dark axion-portal 
study~\cite{Kaneta:2016wvf,Kaneta:2017wfh,Hochberg:2018rjs,
deNiverville:2019xsx,deNiverville:2018hrc,Daido:2019tbm}.

\section{NA64$\mu$ experiment for probing $(g-2)_\mu $ anomaly.
\label{NA64muSect}}
The NA64$\mu$ is upcoming experimental facility at CERN 
SPS~\cite{Gninenko:2014pea,Gninenko:2018tlp,Gninenko:2016ziu,NA64muProposal}, 
which aims to  examine light hidden sector particles weakly coupled to 
muons. It will utilize a muon beam at CERN SPS to search for missing 
energy signatures in the bremsstrahlung process on the active target, 
$\mu N\to \mu N E_{miss}$. That process can be associated with sub-GeV 
hidden vector boson $V$ invisibly decaying into light dark 
matter particles, $V\to \chi\chi$, or neutrinos, $V\to \bar{\nu}\nu$. 
That vector particle is referred to $Z'$-boson, which interacts     
mainly with $L_\mu-L_\tau$ currents of SM. In addition, it can serve a 
sub-GeV vector mediator between SM and  
DM sector due to the mechanism of relic DM 
abundance~\cite{Gninenko:2001hx,Chen:2018vkr,Kahn:2018cqs,
Berlin:2018sjs,Gninenko:2019qiv}. We note however that there are several 
other well motivated scenarios of $Z'$ boson 
which are based on hidden abelian 
symmetries, say $U(1)_{B-L}$ or 
$U(1)_{B-3L_e}$ (for recent review see, 
e.~g.~Refs.~\cite{Berlin:2018bsc,Choudhury:2020cpm}). 

 Furthermore, in Refs.~
\cite{Chen:2017awl,Kahn:2018cqs,Chen:2018vkr}
authors considered a scenarios with muon-specific scalar mediator between 
visible and hidden matter in order to resolve $(g-2)_\mu$ anomaly and 
DM puzzle. One can expect that relevant scalar originates from UV 
completed  models with vector-like fermions and Higgs extended 
sector~\cite{Batell:2016ove,Chen:2015vqy}. 

In Refs.~\cite{Gninenko:2014pea,Gninenko:2018tlp,Gninenko:2016ziu}
 probing of the new dark boson $V$ and hidden scalar $S$ at NA64$\mu$
 was discussed in the light of explanation of the muon magnetic moment anomaly.
In this section we extend the analysis of $V$ and $S$ implication to 
NA64$\mu$~\cite{Gninenko:2014pea,Gninenko:2018tlp,Gninenko:2016ziu}.
In particular, there are two general extensions of the analysis 
discussed in
Ref.~\cite{Gninenko:2014pea,Gninenko:2018tlp,Gninenko:2016ziu}.
 First, we calculate the exact-tree-level production 
cross-section  of hidden neutral boson $S$ and $V$ at NA64$\mu$. That 
analysis is 
based  on the result of Refs.~\cite{Liu:2016mqv,Liu:2017htz} and our 
previous  study~\cite{Gninenko:2017yus} for dark photon production at 
NA64$e$ without using Weizsaecker-Williams approximation in the 
cross-sections of hidden bosons.  Second,
we calculate the expected sensitivity curves of NA64$\mu$ for muon-specific couplings 
of sub-GeV  vector and scalar hidden particles  
$L\supset g_S^\mu S\bar{\mu} \mu + g_V^\mu V_\nu \bar{\mu} \gamma^\nu \mu$ 
by using {\tt GEANT4} Monte-Carlo (MC) simulation.

\begin{figure*}[tbh!]
\begin{center}
\includegraphics[width=0.45\textwidth]{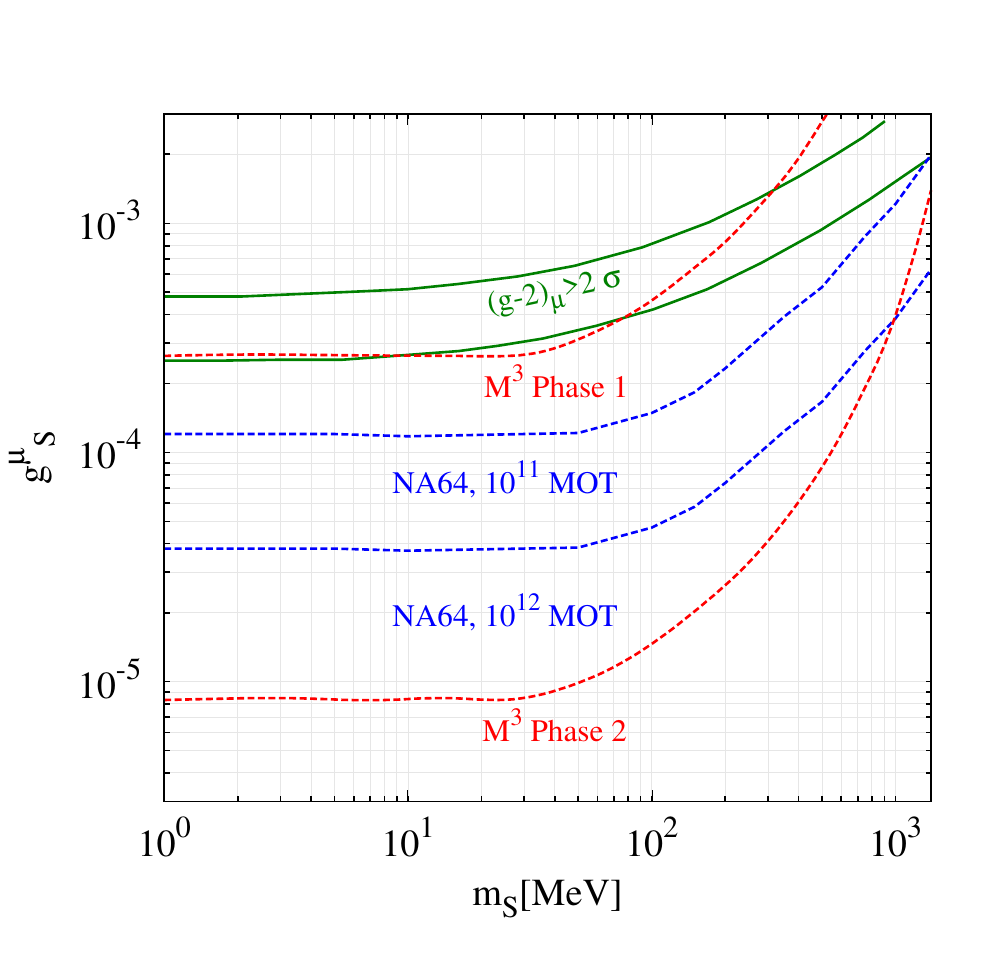}
\includegraphics[width=0.45\textwidth]{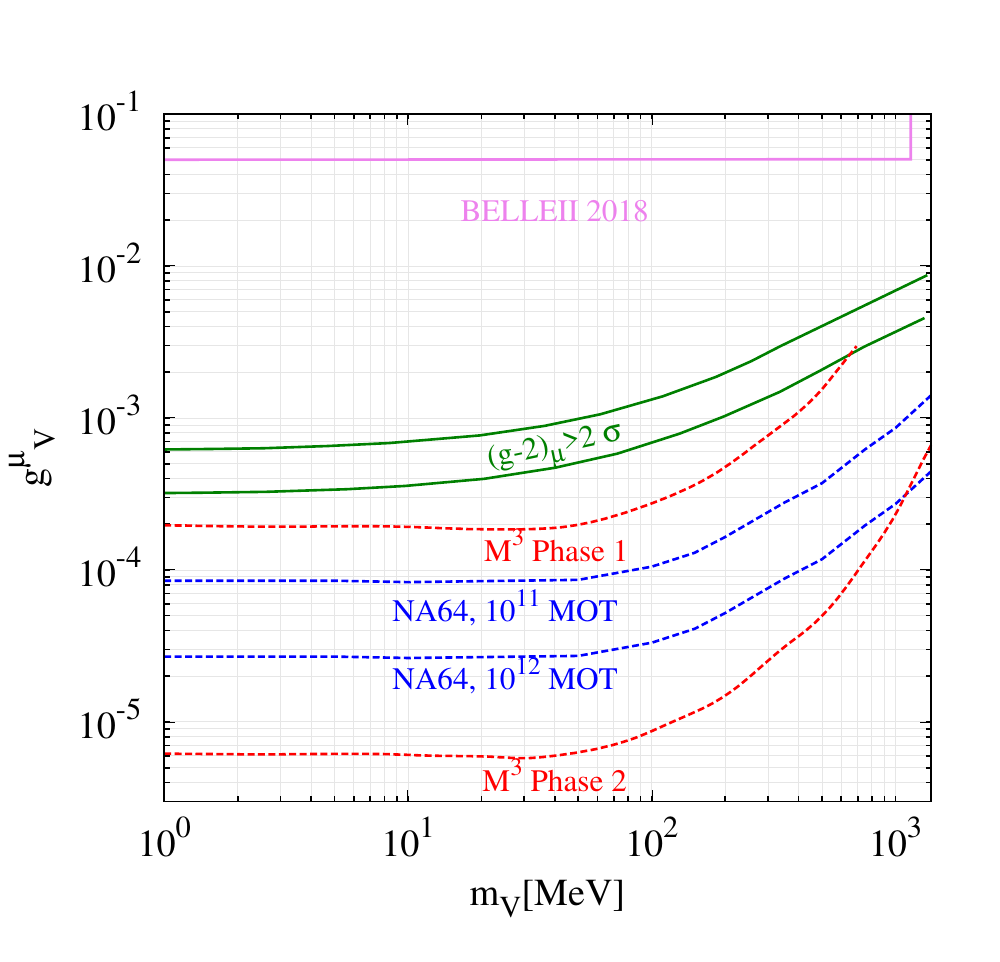}
\label{g-2anomalyScalVecCoupl}
\end{center}
\caption{Left Plot: Dashed blue lines are Dark Scalar expected sensitivities 
of NA64$\mu$ for MOT$=10^{11}$ and MOT$=10^{12}$. 
Right Plot: Dashed blue lines are Dark Vector expected sensitivities of 
NA64$\mu$ for MOT$=10^{11}$ and MOT$=10^{12}$. Corresponding dashed red 
lines are expected limits of $M^3$ experiment~\cite{Kahn:2018cqs} 
for Phase 1 and Phase 2. Pink line shows recent constraints of {\tt BELLEII}
experiment~\cite{Adachi:2019otg} from data collected in 2018. Green 
lines represent the bounds which correspond to resolving of 
$(g-2)_\mu$ anomaly at $2\sigma$ level for both scalar and vector particles. In particular, we use the following inequalities, 
$|\delta a_{V(S)}^\mu - \delta a_\mu^{c.}| < 2 \sigma_{\delta a_\mu}$, where $\delta a_\mu^{c.}=27.9\times 10^{-10}$ and $\sigma_{\delta a_\mu}=7.6\times 10^{-10}$ are taken from Ref.~\cite{Aoyama:2020ynm}.} 
\label{DSgm2PrPzzl}
\end{figure*}

In Fig.~\ref{DSgm2PrPzzl} the expected limits of NA64$\mu$ detector are shown 
for hidden Scalar and Vector boson, we also set benchmark assumption, 
$g_P^\mu=g_A^\mu=0$, such that  pseudoscalar 
and axial vector coupling admixtures don't contribute to $(g-2)_\mu$ anomaly. 
The expected sensitivity of NA64$\mu$ was calculated by using {\tt GEANT4} 
MC simulation of missing energy signal of $E_0=100$~GeV muon scattering on target 
with heavy nuclei $\mu N \to \mu N S(V)$. 
The number of produced light bosons can be approximated as follows
\eq  
 N_{S(V)} \simeq  \mbox{MOT} \times \frac{\rho N_{A}}{A} \times L_T 
 \times \sigma_{S(V)},
\en 
where MOT is a number of muons accumulated on target, $A$ is a atomic weight of 
target medium, $N_{A}$ is Avogadro's number, $\rho$ designates the target density,  
$L_T\simeq 40 X_0$ is a typical distance that are passed by muon before 
producing $S(V)$ with the energy of $E_{S(V)} \gtrsim E_0/2$ 
in the active lead target of NA64$\mu$ ($X_0 \simeq 0.5$ cm), $\sigma_{S(V)}$ 
is a total exact-tree-level production cross-section of light bosons
(for details see, e.~g.~Ref.~\cite{Liu:2016mqv,Liu:2017htz,Gninenko:2017yus}).

 For $m_{S(V)} \lesssim M_\mu$  
that production rate can be approximated in bremsstrahlung-like limit as  
$\sigma_{S(V)} \sim (g^\mu_{S(V)})^2/M_\mu^2$.  Which implies that relevant 
sensitivity curves in Fig.~\ref{g-2anomalyScalVecCoupl} have a plateau  
in the light mass region. In Fig.~\ref{g-2anomalyScalVecCoupl}  
we require $N_{S(V)} > 2.3$, which corresponds to $90\%$ CL exclusion bound 
on  $g^\mu_{S(V)}$ coupling for the background free case. In particular,  
a preliminary hadron contamination analysis and study of the detector  
hermiticity with muon beam~\cite{NA64muProposal} show that total background 
to be at the level $\lesssim 10^{-12}$. It is worth mentioning that muon energy 
loses in the lead target can be neglected~\cite{Chen:2017awl}, since the muon energy 
attenuation is small for typical beam energy, 
$\langle dE_\mu/dz  \rangle \simeq 12.7\cdot 10^{-3}$ GeV/cm.

In the NA64$\mu$ experiment one assumes to utilize two, upstream and downstream, 
magnetic spectrometers. These spectrometers, will provide a precise measurements of 
initial and final muon energies~\cite{NA64muProposal}. 
We suppose that $S(V)$ being produced  by muons in the target escapes  
the NA64$\mu$ detector without interaction decaying invisibly into DM particles. 

Indeed, let us estimate the absorption length of light scalar and vector
in the medium of the NA64 detector. Hidden bosons  should be sterile 
enough to avoid energy deposition in ECAL and HCAL due to their absorption. 
In high energy limit, $E_{V(S)} \gg 10$ GeV, the number  QCD resonances of 
$\sim 100$ MeV produced due to the boson absorption  by the protons will 
be negligible. So that, the leading process of boson attenuation in 
calorimeters is inelastic scattering  $ V(S) + p \to p +$jet. For relatively light bosons, 
$m_{V(S)} \lesssim 100$ MeV, one can estimate the absorption 
cross-section  in high-energy regime, $E_{V(S)} = E_{H} \gg m_p$, as 
$\sigma_{abs} \simeq Z \alpha_s (g^{p}_{H})^2/(4 s)$, 
where $s \simeq 2 E_H m_p$ is a  center of mass energy squared.
Here we denote $H=(V,S)$.  
For estimate we take $g^p_{V(S)} \simeq 10^{-4}$  as a typical 
benchmark  coupling of  hidden bosons to proton 
(see, e.~g.~Tab.~\ref{tab:limitsX17}  below).
 Therefore for  $\alpha_s(M_Z)=0.12$ one has  
\eq 
\sigma_{abs} \simeq 1.2 \cdot 10^{-31} \, \mbox{cm}^2\, (g_H^p)^2 Z  
\left( \frac{E_H}{100\mbox{GeV}} \right)^{-1}\,. 
\en
Now we can estimate the absorption length as 
$\lambda_{abs} \simeq (n  \, \sigma_{abs})^{-1}$, where 
$n= \rho N_A /A$ 
is a typical density number of atoms in the medium of the target.
For iron medium of hadronic calorimeter, $\rho = 7.87$~g/cm$^3$,   
$A = 56$~g/mole, $Z= 26$, $E_{H} = 100$ GeV, $g_H^p=10^{-4}$, one obtains
$ n^{Fe} \simeq 2.2 \cdot 10^{24}$~cm$^{-3}$ and 
$\lambda_{abs} \simeq 5 \cdot 10^{13}$~cm.
This means that light bosons produced by muon beam in target 
will pass the hadronic calorimeter module without energy deposition.

\section{$^8$Be anomaly constraints
\label{8BeSect}}

It is worth mentioning that nucleon terms in 
Lagrangian~(\ref{NuclAndLept}) can be referred to hadron-$X17$ boson  
couplings~\cite{Feng:2016jff,Feng:2016ysn} for the case of parity-violating 
interaction~\cite{Kahn:2016vjr}. In particular, authors 
of~\cite{Feng:2016jff,Feng:2016ysn} provide a rough estimate of $P$-even
hadronic couplings  of $X17$ boson as 
$|f^p_A/e|\simeq|g^p_V /e|\lesssim 1.2 \times 10^{-3}$ and  
$|f^n_A/e|\simeq |g^n_V/e|\lesssim (2-10)\times 10^{-3}$ from null result of 
$\pi^0 \to \gamma (X17\to e^+e^-)$ decay 
at NA48/2~\cite{Raggi:2015noa,Batley:2015lha} and best fit of $X17$ decay 
in the ATOMKI experiment~\cite{Krasznahorkay:2015iga}. For $P$-odd hadronic 
couplings of $X17$ vector boson one can expect them to be proportional to 
quark axial couplings $g^{n(p)}_A \simeq f^{n(p)}_V \sim g_q^A$ in a 
manner of Ref.~\cite{Kozaczuk:2016nma}. Namely, a comprehensive analysis 
of~\cite{Kozaczuk:2016nma} for both, enhanced isoscalar, 
$^8\mbox{Be}^{*\prime}(J^P=1^+; T=0)\to ^8\mbox{Be}^*(J^P=0^+; T=0)+X17$, 
and suppressed isovector, 
$^8\mbox{Be}^{*}(J^P=1^+; T=1) \to\, ^8\mbox{Be}^*(J^P=0^+; T=0)+X17$,  
nuclear transitions implies a conservative bounds 
$|g^{n(p)}_A|\simeq|f^{n(p)}_V |\lesssim 10^{-5}-10^{-4}$. 
The hadronic terms in the Lagrangian~(\ref{NuclAndLept}) involving hidden 
scalar and pseudoscalar particles can be originated from extended Higgs 
sector of SM~\cite{Ellwanger:2009dp}. In particular, light pseudoscalar can
be a valid candidate  for $^8$Be anomaly 
explanation~\cite{Ellwanger:2016wfe}. The relevant Lagrangian reads
\eq 
\mathcal{L} \supset \sum_{q=u,d} \xi_p \frac{m_q}{v} P \bar{q} i 
\gamma_5 q,
\label{PseudRedCoupl}
\en  
where $v=246$ GeV is the Higgs vacuum expectation value. 
This implies~\cite{Ellwanger:2016wfe} that the resulting 
Yukawa-like couplings of $P$ to  up and down type quarks 
are $\xi_u\simeq \xi_d \simeq 0.3$, with $\xi_u$ and $\xi_d$ 
being a linear combination of nucleus couplings, such that  
\eq
g_P^p\simeq \frac{M_p}{v}(-0.40 \xi_u -1.71 \xi_d), 
\en
\eq 
g_P^n\simeq \frac{M_n}{v}(-0.40 \xi_u +0.85 \xi_d)\,. 
\en 
Therefore, one has conservative limits, 
$|g_P^p| \lesssim 2.5 \times 10^{-3}$ and 
$|g_P^n| \lesssim 5.5 \times 10^{-4}$, which, however depend  
on nuclear shell model of isospin transition~\cite{Ellwanger:2016wfe}.
We note, that Lagrangian~(\ref{PseudRedCoupl}) doesn't respect 
gauge symmetry of SM unbroken gauge group, and therefore can be 
considered as effective interaction of UV completed 
model~\cite{Domingo:2016yih}. 

Now let us consider $^8$Be constraints for light hidden boson from  lepton sector,  
which is described by the second  term  in the Lagrangian~(\ref{NuclAndLept}). 
A numerous well motivated scenarios~\cite{Gu:2016ege,Chen:2016dhm,%
Liang:2016ffe,Jia:2016uxs,Kitahara:2016zyb,Chen:2016tdz,%
Seto:2016pks,Neves:2016ugb,Chiang:2016cyf,Krasnikov:2017dmg,%
Neves:2017rcn,Zhu:2017moa,Neves:2019zog,Pulice:2019xel,Nam:2019osu}  
have been suggested recently for explaining the ATOMKI $e^+e^-$ anomaly, 
which involve neutral vector boson interacting with leptons.
That vector particle decays visibly via $e^+e^-$ pair, 
with $\mbox{Br}(V\to e^+e^-)\simeq 1$, since 
its mass doesn't exceed the masses of any hadronic states. 
The dominant constraints on vector coupling to electron come from 
NA48/2 data on $ \pi^0\to \gamma V(V\to e^+e-)$ decay and from NA64e data 
on $eN\to eNV (V\to e^+e^-)$ bremsstrahlung  $e^+e^-$ pair emission. 
In particular, NA48/2 experimental facility provides best upper limit 
on $X17(J^P=1^+)$ mixing with electrons, 
$\mathcal{L} \supset g_V^e V_\mu \bar{e}\gamma^\mu e$, such that 
the allowed values of coupling are 
$g_V^e/e \lesssim  1.4 \times 10^{-3}$ at $90 \%$ CL.  
NA64e experiment has been recently set the lower limit 
on the relevant coupling at $90 \%$~CL~\cite{Banerjee:2019hmi}. 
Therefore, the existence of $X17$ vector boson favors the following 
values of electron mixing $g_V^e/e \gtrsim  6.8 \times 10^{-4}$. 
The former bound can be rescaled for the case of axial-vector coupling 
admixture, $\mathcal{L} \supset V_\mu \bar{e}\gamma^\mu(g_V^e  
+ \gamma_5 g_A^e) e$, as  $\sqrt{(g_V^e)^2+(g_A^e)^2}/e\gtrsim  
6.8 \times 10^{-4}$. 
 
\begin{table*} 
\begin{center}
\caption{Favored couplings for $X17(J^P=1^\pm, 0^-)$} 

\vspace*{.15cm}

\def\arraystretch{1.5}
\begin{tabular}{c|c|c|c}
\hline
\hline
 \ \ Coupling 
 \ \ & \ \ Neutron \ \ & \ \ Proton \ \ & \ \ Electron \ \
\\
\hline
\hline
$g_P/e=f_S/e$
& $ \lesssim  1.8 \times 10^{-3}$\, \mbox{from Ref.~\cite{Ellwanger:2016wfe}} 
& $\lesssim  8.3 \times 10^{-3}$\, \mbox{from Ref.~\cite{Ellwanger:2016wfe}} 
& $\gtrsim 3.0 \times 10^{-5}$\, \mbox{from Refs.~\cite{Ellwanger:2016wfe,Andreas:2010ms}}  
\\
$g_V/e=f_A/e$ 
& $ \lesssim  (2-10) \times 10^{-3}$\, \mbox{from Ref.~\cite{Feng:2016jff,Feng:2016ysn}} 
& $ \lesssim  1.2 \times 10^{-3}$\, \mbox{from Ref.~\cite{Feng:2016jff,Feng:2016ysn}} 
& $\lesssim 1.4 \times 10^{-3}$ \, \mbox{from Ref.~\cite{Batley:2015lha}}\,,
\\
&
&
& $\gtrsim 6.8 \times 10^{-4}$ \, \mbox{from Ref.~\cite{Banerjee:2019hmi}} 
\\ 
$g_A/e=f_V/e$ 
& $\lesssim  3.3\times(10^{-5} - 10^{-4})  $\, \mbox{from Ref.~\cite{Kozaczuk:2016nma}} 
& $\lesssim  3.3\times(10^{-5} -  10^{-4})  $\, \mbox{from Ref.~\cite{Kozaczuk:2016nma}} 
&  $\gtrsim 6.8 \times 10^{-4}$ \, \mbox{from Ref.~\cite{Banerjee:2019hmi}}
\\ 
\hline 
\hline
\end{tabular}
\label{tab:limitsX17}
\end{center}
\end{table*}

It is worth mentioning that one can estimate the projected sensitivity of 
the NA64 to probe pseudo-scalar  particle $X17(J^P=0^-)$ which decays
visibly to electron-positron pair, Br$(P \to e^+e^-)=1$. 
The authors of Ref.~\cite{Ellwanger:2016wfe} provide the following limits 
for reduced  Higgs-like coupling of $X17(J^P=0^-)$ boson with electrons 
$\mathcal{L} \supset \xi^e_P \frac{M_e}{v} P \bar{e}i \gamma_5 e$
\eq 
 \xi_P^e \gtrsim  4.5   
\label{XiLimFromAndreaPaper}
\en 
which are favored by experimental data from electron and proton beam-dump 
facilities~\cite{Andreas:2010ms} for $m_P \gtrsim 17$ MeV. The relevant 
limits $C_{Aff}=\xi_P^e$  are  shown in Fig.~(4) of 
Ref.~\cite{Andreas:2010ms}. These bounds can be transferred to the 
electron's coupling in terms of Lagrangian~(\ref{NuclAndLept}) as follows  
$
g_P^e/e \gtrsim 3.0\times 10^{-5},     
$
from {\tt CHARM} data~\cite{Andreas:2010ms}. 
It is instructive to compare $g_P^e/e$ limits with the corresponding 
bounds of vector boson   
$X17(J^P=1^+)$, that has the allowed couplings in the range 
$6.8\times 10^{-4}\lesssim g_V^e/e \lesssim 1.4\times 10^{-3}$.
To summarize results, let us estimate the   
lifetimes~\cite{Liu:2016mqv,Liu:2017htz} of $P$ and $V$ which 
have not been  experimentally excluded yet. In particular, 
one has the following  constraints from various experiments 
\begin{itemize}
\item {\tt CHARM}~\cite{Andreas:2010ms}:  $\tau_P \lesssim 1.2\times 10^{-11}$~s,  $g_P^e/e \gtrsim 3.0\times 10^{-5},$
\item NA48/2~\cite{Batley:2015lha}: $\tau_V \gtrsim 8.3\times 10^{-15}$~s, $g_V^e/e \lesssim  1.4 \times 10^{-3},$
\item NA64e~\cite{Banerjee:2019hmi}: $\tau_V \lesssim 3.5\times 10^{-14}$s, $g_V^e/e \gtrsim 6.8\times 10^{-4}$.
\end{itemize}
Note, that corresponding bounds on lifetimes of $X17$ boson are in 
agreement with the estimate $\tau_{X17} \lesssim  10^{-10}$~s 
from ATOMKI data~\cite{Feng:2016jff} as expected for both 
pseudoscalar and vector realizations of $X17$. 
 In particular, authors 
of Ref.~\cite{Feng:2016jff} require that $X17$ decays
into $e^+e^-$ within $L \lesssim 1$~cm, where $L = c  \tau_{X}  \beta_X \gamma_{X}$, here $\gamma_{X} \simeq 1.06$ and $\beta_X=0.35$.  

We point out that the NA64e experiment has an excellent prospect for 
probing of  $X17(J^P=0^-)$, since it will decay mostly 
into $e^+e^-$  within the fiducial volume of the NA64e ($L_{fid} \sim 7- 10$ m) 
due to large boost factor 
$E_P/m_P\simeq 6\times 10^3$, with typical decay length of 
$L_{dec}^P \simeq 14$~m. We note however that our  estimate is conservative, 
therefore one  should perform a comprehensive Monte-Carlo simulation for 
the flux and  spectra of  hidden pseudo-scalars produced in the target by 
primary  electrons, $eN \to eN P (P\to e^+e^-)$. That investigation will 
take into account realistic response and efficiency of the NA64e detector. 
We leave that task for future analysis~\cite{GninenkoInPrepar}. 
In Tab.~\ref{tab:limitsX17} we summarize current limits on $X17$ couplings.

\section{Combined contribution of light bosons to the proton radius
\label{c}}

\begin{figure}[t!]
\begin{center}
\includegraphics[scale=0.8]{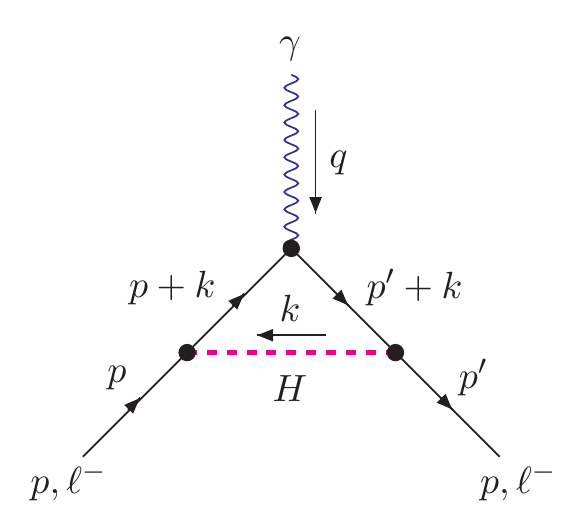}
\end{center}
\caption{Vertex correction to anomalous magnetic moment
of proton and charged lepton due to
exchange of new particles. }
\label{fig:diag}
\end{figure} 
 
In this section we consider the problem of the proton charge radius. 
In particular, we discuss direct contribution of these light bosons into 
the proton radius via the  charge Sachs form factor $G_E^P(q^2)$. 
As we stressed in Sec.~\ref{IntroSec},  
this possibility is quite interesting in the connection 
to planned precise measurement of the proton charge radius from analysis 
of the elastic muon-proton scattering. 
$P$-even electromagnetic vertex function is defined
for incoming photon as
\eq 
\hspace*{-.2cm}
M_{\rm inv}^P = \bar u(p') \, \biggl[ \gamma^\mu \, F_1(q^2) \,+\,
  \frac{i}{2M_p} \, \sigma^{\mu\nu} q_\nu \, F_2(q^2) \biggr]  u(p) \,. 
\en
Here $F_1$ and $F_2$ are the Dirac and Pauli form factors;
$q^2 = -Q^2$. For minimal coupling of photon with proton and charged leptons
\eq
{\cal L}_{\rm em; m} = e A_\mu \, [\bar p \gamma^\mu p -
\bar\ell \gamma^\mu \ell].
\en

It is interesting to look at the relative contribution of new hidden 
particles to both proton charge radius and muon $(g-2)_\mu$ ratio 
anomaly. We do the direct estimate of the contributions of new particles 
into the proton charge radius. The proton charge radius is defined as
\eq 
\langle r_p^E \rangle^2 = - 6 [G_E^P(0)]' = - 6 [F_1^p(0)]' +
\frac{3}{2 M_p^2} F_2^p(0)\,,
\en
where $F_1^p$ and $F_2^p$ are the Dirac and Pauli
electromagnetic form factors of the proton, respectively; 
$[F(0)]'$ means the derivative with respect to $Q^2$ at $Q^2=0$. 
Here $F_2^p(0) = \kappa_p$ is the proton anomalous magnetic moment.
In particular, using the mass value $m_V = 16.7$ MeV  
of the hypothetic $X17$ vector particle observed 
in the ATOMKI experiment~\cite{Krasznahorkay:2019lyl} we 
get the following leading (logarithmic) 
contribution to the charge proton radius: 
\eq
\langle \delta r_p^E \rangle^2   
\simeq 0.014 \, h_r^{(1)} \ {\rm fm}^2 \,,
\en
where 
\eq 
h_r^{(1)} = 
(g_V^p)^2 - (g_A^p)^2 + (f_A^p)^2 - (f_V^p)^2 \,,
\en
is the combination of couplings of vector and axial 
vectors with proton (see Appendix). Let us estimate that contribution
for benchmark couplings shown in Tab.~\ref{tab:limitsX17} for $X17$ boson. 
In particular, 
$h_r^{(1)} = 2(g_V^p)^2 - 2(g_A^p)^2 \simeq 2.6 \times 10^{-7} $, that 
yields $\langle \delta r_p^E \rangle \simeq 6 \times 10^{-5}$ fm. 
Therefore we conclude, that current information on new particles suggests 
that their contribution to the charge radius of proton 
is negligible. We note, that the small impact of 
BSM  effects on proton charge radius was discussed originally in 
Ref.~\cite{Karshenboim:2014tka}.

\section{Constraints on couplings of new particles using                      
data on electric dipole moments of leptons and neutron
\label{EDMSect}}

\begin{figure}[tbh!]
\begin{center}
\includegraphics[width=0.5\textwidth]{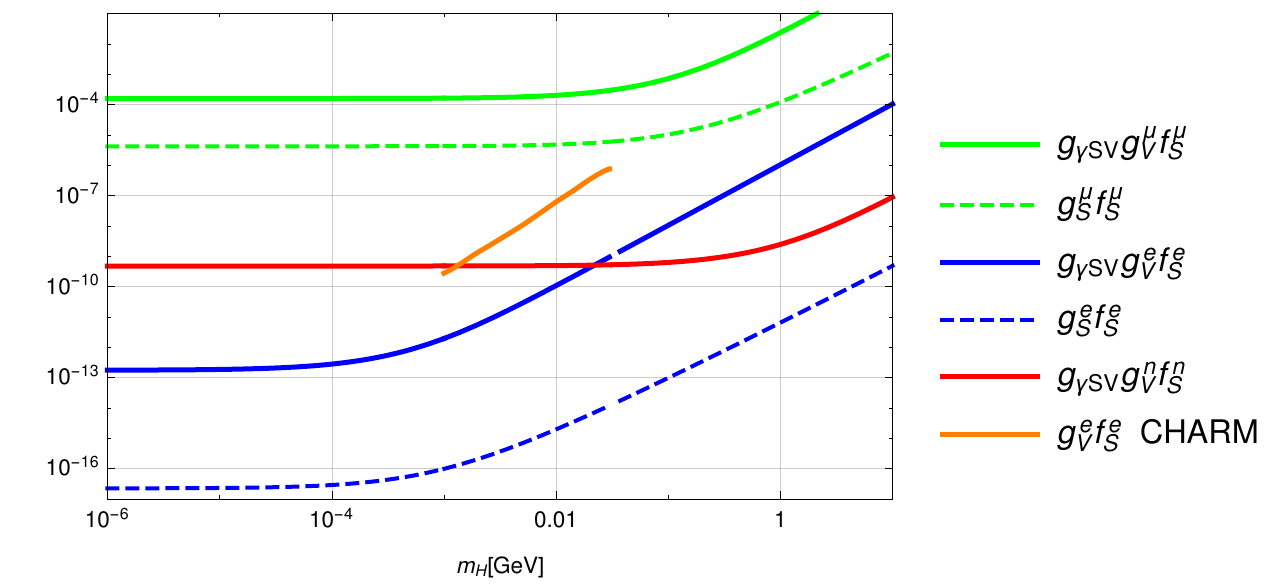}
\end{center}
\caption{$90\%$ CL constrains on coupling combinations 
from EDM of SM fermions. Solid (dashed) green line shows bound on $ g_{\gamma SV} g_{V}^\mu f^\mu_S$ $(g_{S}^\mu 
f^\mu_S)$ coupling from EDM of muon. Solid (dashed) blue line shows bound on $ g_{\gamma SV} g_{V}^e f^e_S$ $(g_{S}^e 
f^e_S)$ coupling from EDM of electron. Solid red line 
shows bound on $ g_{\gamma SV} g_{V}^n f^n_S$ 
 coupling from EDM of neutron. Solid orange line represents combined limits on $g^e_V f_S^e$ from 
electron EDM and {\tt CHARM} bounds.} 
\label{EDM90bounds}
\end{figure}

\begin{figure}[htb!]
\begin{center}
\includegraphics[scale=0.55]{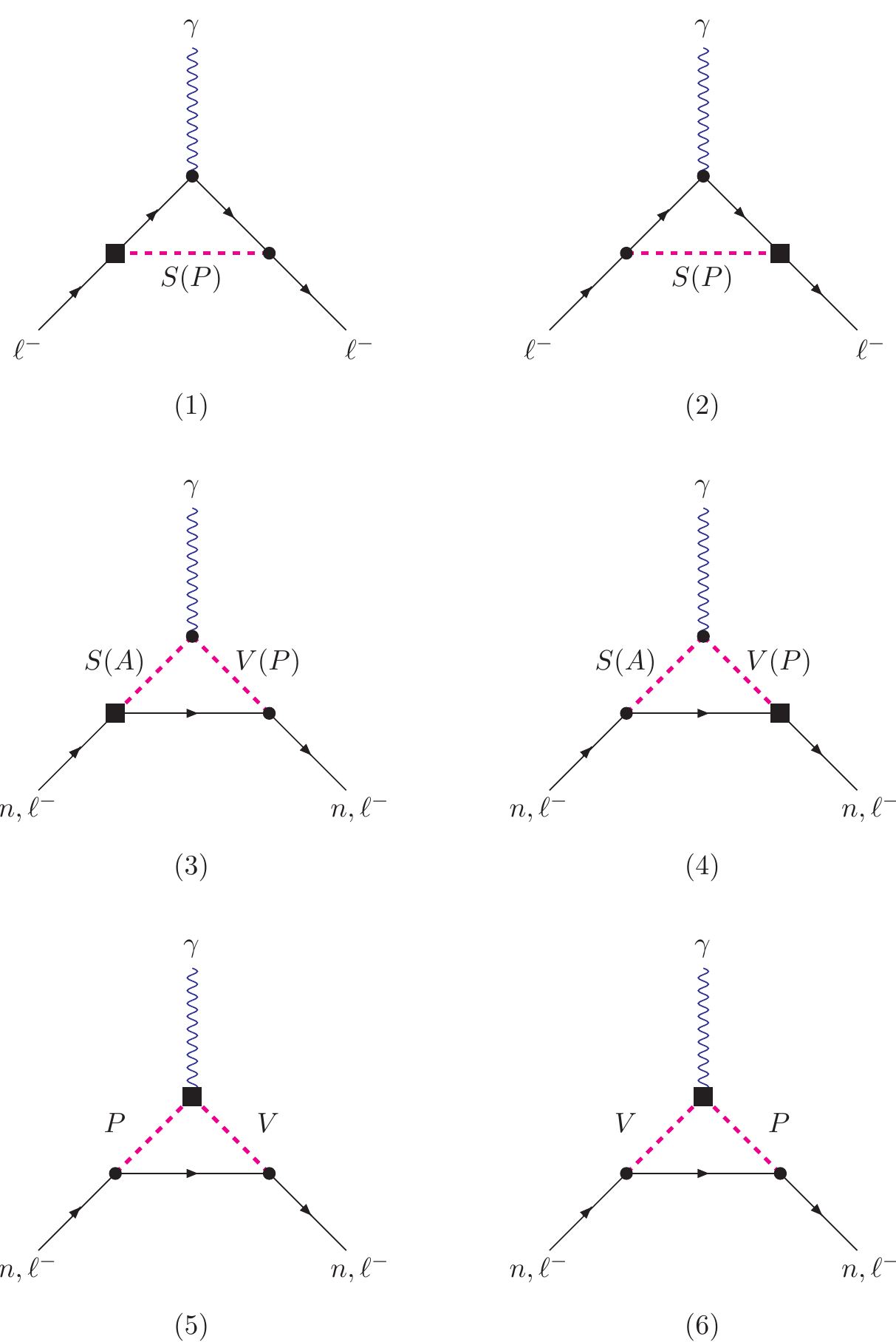}
\end{center}
\caption{Diagrams describing contribution of new particles to
electric dipole moments of neutron and leptons. The square boxes denote 
the $P$-odd vertices, round vertices are $P$-even couplings.}
\label{fig:diag2}
\end{figure}

In this section we derive the constraints on the combinations
of $P$-even and $P$-odd couplings of new particles using
data on electric dipole moments (EDM) of leptons and neutron.
The contributions of new particles to EDMs are described
by the diagram in Fig.~\ref{fig:diag2} where squared vertex
is $P$-odd
and round vertex is $P$-even  coupling with leptons (neutron).
The EDM of spin-$\frac{1}{2}$ fermion $\psi$ (neutron or leptons) 
is defined as $d^E=D_E(0)$, where $D_E(q^2)$ is the relativistic 
electric dipole form factor extracted from full electromagnetic 
vertex function of corresponding fermion~\cite{EDM}: 
\eq 
M_{\rm inv} &=& \bar u_\psi(p_2)\,\Gamma^\mu(p_1,p_2)\,u_\psi(p_1)\,,
\nonumber\\
\Gamma^\mu(p_1,p_2) &=& 
- \sigma^{\mu\nu}q_{\nu} \gamma^5 \, D_E(q^2) 
\,+\, \ldots 
\label{vertex}
\en
The contributions of individual diagrams in Fig.~\ref{fig:diag2} 
are given in Appendix. 

\begin{table*}
\caption{Upper limits on couplings of new particles 
from data on EDMs of electron, muon, and neutron} 

\def\arraystretch{1.5}
\begin{tabular}{c|c|c|c}
\hline
\hline
 \ \ Coupling combination 
 \ \ & \ \ Electron \ \ & \ \ Muon \ \ & \ \ Neutron \ \
\\
\hline
\hline
$|g_P \, f_P|=|g_S \, f_S|$, $m_H\ll M_\psi$  
& $< 2.7 \times 10^{-17}$ 
& $< (0.4-3.8) \times 10^{-5}$ 
& $-$ 
\\
$|f_{\gamma PV} \, g_P \, g_V|=|g_{\gamma PA} \, 
g_P \, f_A|=|g_{\gamma SV} \, g_V \, f_S|$, $m_H\ll M_\psi$  
& $< 1.7 \times 10^{-13}$ 
& $< (0.2-1.4) \times 10^{-3}$ 
& $< 4.5 \times 10^{-10}$ 
\\ 
$|f_{\gamma PV}|=|g_{\gamma PA}|=|g_{\gamma SV}|$, $m_H \simeq 16.7$ MeV  
& $ \lesssim 7.7 \times 10^{-2} $
& $-$ 
& $\lesssim 1.5\times 10^{-3}- 3\times 10^{-4}$ 
\\ 
\hline 
\hline
\end{tabular}
\label{tab:limits}
\end{table*}
Using the upper limits/results for the electron, muon, and neutron 
EDMs: 
\eq 
& &|d^E_e| < 1.1 \times 10^{-29} \, e \, \mbox{cm, \quad at \,\, 90 \%\, CL,}\,\, 
\mbox{Ref.~\cite{Andreev:2018ayy}}\,,  
\nonumber\\
& &|d^E_\mu| < 1.8  \times 10^{-19} \, e \, \mbox{cm, \quad at \,\, 95 \%\, CL,}\,\, 
\mbox{Ref.~\cite{Bennett:2008dy}}\,, 
\nonumber\\
& &|d^E_n| < 3.0 \times 10^{-26} \, e \, \mbox{cm, \quad at \,\, 90 \%\, CL,}\,\, 
\mbox{Ref.~\cite{Afach:2015sja}} \,,  
\nonumber
\en 
we get the  upper limits for combinations of couplings of 
new particles, which are displayed in Tab.~\ref{tab:limits}. 
Let us consider several benchmark limits. Namely, for concreteness we set 
to  zero couplings of dimension-five operators~(\ref{AxionPortalCouplings}), 
$g_{\gamma S V}=g_{\gamma PA}=f_{\gamma PV}=0$. That yields the 
following constraints 
 on electron and muon interaction with $P$ and $S$ for light bosons masses
 $m_V=m_P=m_S \ll M_e$ 
\eq 
|d_e/e| \simeq \frac{|g_S^e f_S^e |}{8 \pi^2 M_e}  < 
5.5 \cdot 10^{-16} \,\, \mbox{GeV}^{-1},
\en 
\eq 
|d_\mu/e| = \frac{|g_S^\mu f_S^\mu |}{8 \pi^2 M_\mu} < 
5.0 \times 10^{-7} \,\, \mbox{GeV}^{-1} \,,
\en 
or equivalently $|g_S^e f_S^e |< 2.7 \times 10^{-17}$ 
and $|g_S^\mu f_S^\mu | < 4.2 \times 10^{-6} $. 
In order to avoid interference between diagrams (1)-(4) in Fig.~\ref{fig:diag2}  
we now consider a benchmark point $g_S=g_P=0$.  That implies the following limits 
or the product of vector-specific and pseudo-scalar couplings of leptons
\eq 
|d_e/e| = \frac{|g_{\gamma S V} g^e_V f^e_S|}{16 \pi^2 M_p} \frac{1}{2} < 
5.5 \cdot 10^{-16} \, \mbox{GeV}^{-1}, 
\en
\eq
|d_\mu/e| = \frac{|g_{\gamma S V} g^\mu_V f^\mu_S|}{16 \pi^2 M_p} \frac{1}{2} < 
5.0 \times 10^{-7}  \mbox{GeV}^{-1},
\en  
which yield $|g_{\gamma S V} g^e_V f^e_S| < 1.7 \times 10^{-13}$ and 
$|g_{\gamma S V} g^\mu_V f^\mu_S| < 0.2 \times 10^{-3}$. For relatively 
light hidden bosons $m_{A}=m_V=m_P=m_S \ll M_n$ one can also obtain corresponding 
constraint from neutron EDM,
$|g_{\gamma SV} g^n_V f^n_S| < 4.5 \times 10^{-10}$. Heavy bosons $m_{H}\gg M_\psi$ 
yield the limits on coupling products, which are scaled as $ \sim (m_{H}/M_\psi)^2$. 
These bounds are shown in Fig.~\ref{EDM90bounds}. One can see 
from Fig.~\ref{EDM90bounds}, that the most stringent constraints on couplings come 
from electron EDM bounds for $m_H \ll M_e$. Moreover, for the benchmark 
values of electron coupling with vector,  
$g_V^e/e \simeq 1.4 \times 10^{-3}$, and scalar, 
$f_S^e/e \simeq 3.0 \times 10^{-5}$, one can also estimate the bound on 
$g_{\gamma SV}$  that is favored by $X17$-boson existence. In particular, 
for $m_H \simeq 16.7$ MeV, one has $g_{\gamma V S} \lesssim 7.7 \times 10^{-2}$  
from Fig.~\ref{EDM90bounds}. Corresponding bound from 
neutron EDM yields $g_{\gamma V S} \lesssim 1.5\times 10^{-3}- 3\times 10^{-4}$ 
for $g^n_V/e \simeq (2 -10)\times 10^{-3}$ and $f_S^n/e\simeq 1.8\times 10^{-3}$
provided in Tab.~\ref{tab:limitsX17}. 
Here we expect naively that $X17$ is  admixture of vector and pseudo-scalar 
states which have dark axion portal coupling as 
in Ref.~\cite{deNiverville:2019xsx,deNiverville:2018hrc} 
$\mathcal{L}\supset \frac{1}{2} a\, G_{a\gamma \gamma'} F_{\mu\nu}F_{\mu\nu}'$. 
In particular, one can relate corresponding values 
of  $G_{a \gamma \gamma'}$ and $g_{\gamma S V}$  as follows, 
$G_{a\gamma\gamma'}=e g_{\gamma SV}/(2 M_{p})$. That implies conservative 
bound on dark axion portal interaction of $X17$ states 
$G_{a\gamma \gamma'} \lesssim 2.5\times 10^{-4} - 5\times 10^{-5}\, \mbox{GeV}^{-1}$ 
for $m_a=m_{\gamma'}\simeq 16.7$ MeV. We note that our latter rough estimate 
is referred to the model, which   incorporates consistently both 
$X17(J^P=0^-)$ and  $X17(J^P=1^+)$ states for $^8$Be anomaly explanation. 
The development of that scenario however is beyond the scope of the 
present paper.  Besides, we want to point out that proposed sensitivity for a future
measurement of the proton EDM and indirect limit to neutron EDM which 
the JEDI  Collaboration \cite{Abusaif:2019gry} plans to obtain at 
level of $\sim 10^{-29}$ can
 receive more stringent limit for the couplings by a factor  
 $10^{-3}$. We note that relevant limits for the combinations
 of couplings were set recently  in Refs.~  \cite{Stadnik:2017hpa,Dzuba:2018anu} for the 
 axion-like particle in the wide range of its masses 
$10^{-8}\, \mbox{eV}\, \lesssim  m_a \lesssim  10^{12}\, \mbox{eV}\
$.

It is instructive to obtain constraint on $g_V^e f_S^e$ coupling from 
combined limit on electron EDM and {\tt CHARM} experimental  
bounds for dark axion portal interaction $G_{a\gamma \gamma'}$  presented 
in Ref.~\cite{deNiverville:2019xsx,deNiverville:2018hrc}.   
The authors of Ref.~~\cite{deNiverville:2019xsx,deNiverville:2018hrc}  
have been set severe upper limit on $G_{a\gamma\gamma'}$ assuming   
null result of {\tt CHARM} experiment to observe   
$\gamma' \to a \gamma$  decay within regarding fiducial volume.   
The latter  implies $m_{\gamma'} \gg m_{a}$, thus contribution of   
$\gamma'$ and $a$ into $d_e/e$ in that mass range reads as follows 
\eq 
   d_e/e = \frac{G_{a \gamma \gamma'} f_a^e g_{\gamma'}^e}{8\pi^2} 
\,   J\left(\frac{m_{\gamma'}}{m_e}, 0 \right)\,.
\en 
Here we use the notations of 
Ref.~\cite{deNiverville:2019xsx,deNiverville:2018hrc} 
denoting indices as $a=S$ and $\gamma'=V$ for axion-like and dark-photon 
particles respectively, the function $ J\left(m_{\gamma'}/m_e, 0 \right)$ 
is given by Eq.~(\ref{Jmutau}) in Appendix. 
In particular, for $m_\gamma' \gg M_e$ one has 
\eq
|g_{\gamma'}^e f_a^e|  < 1.3\times 10^{-13} 
\left(\frac{G_{a\gamma\gamma'}}{\mbox{GeV}^{-1}}\right)^{-1}
\, \frac{m_{\gamma'}^2/M_e^2}{\log(m_{\gamma'}^2/M_e^2)} \,,
\en
which yields $10^{-10 }\lesssim|g_{\gamma'}^e g_{a}^e | \lesssim 10^{-6}$  
for the masses in the range  
$1\, \mbox{MeV}\lesssim m_{\gamma'} \lesssim 30 \, \mbox{MeV}$ 
from {\tt CHARM} experimental constraints in Fig.~3 of 
Ref.~\cite{deNiverville:2019xsx}. We show corresponding limit in 
Fig.~\ref{EDM90bounds} by solid orange line. 
 
\section{Summary
\label{summary}}

In this paper we discuss phenomenological aspects of new scalar, pseudoscalar, 
vector and axial particles coupled to fermions (nucleons and leptons), which 
could give contributions to proton charge radius and $(g-2)_\mu$ ratio, 
$^8$Be anomaly and EDM of fermions. 
The main conclusions of this paper are:

\begin{itemize}
\item We estimate sensitivity of NA64$\mu$ muon active target experiment 
to probe sub-GeV Vector and Scalar mediator of DM by using comprehensive
{\tt GEANT4} MC simulation. These bosons can possibly
 explain $(g-2)_\mu$ anomaly. In case of NA64$\mu$ null result of observing 
 muon missing energy events associated with hidden vector and scalar particles, 
 $\mu N \to \mu N S(V)$, one can exclude new sub-GeV bosons as interpretation of $(g-2)_\mu$ 
 anomaly. 
\item We summarize existing constraints on $^8$Be anomaly 
for  hidden $X17(J^P=0^-,1^\pm)$ bosons. We  estimate contribution of 
these particles to
proton charge radius by direct calculation of Sachs form factors. It turns out that the 
resulting contribution is negligible. 
\item We also set constraints on   couplings of dimension-five 
operators for light hidden bosons
which can contribute to EDM of SM fermions. That novel EDM analysis is motivated by 
dark axion portal study, which involves axion-photon-dark-photon couplings. 

\end{itemize}
 
\section*{Acknowledgments}
 We would like to thank S.~N.~Gninenko, N.~V.~Krasnikov and M.~M.~Kirsanov for many
fruitful discussions. 

The work of V.~E.~L. was funded by
``Verbundprojekt 05P2018 - Ausbau von ALICE                                                
am LHC: Jets und partonische Struktur von Kernen''
(F\"orderkennzeichen No. 05P18VTCA1),
``Verbundprojekt 05A2017 - CRESST-XENON: Direkte Suche nach Dunkler                                                
Materie mit XENON1T/nT und CRESST-III. Teilprojekt 1''
(F\"orderkennzeichen 05A17VTA)'', the Carl Zeiss Foundation under
Project “Kepler Center f\"ur Astro- und Teilchenphysik: Hochsensitive
Nachweistechnik zur Erforschung des unsichtbaren Universums (Gz: 0653-2.8/581/2)”
by ANID PIA/APOYO AFB180002 and by FONDECYT (Chile) under Grant No. 1191103,
by the Tomsk State University Competitiveness Enhancement Program
``Research of Modern Problems of Quantum Field Theory and Condensed Matter Physics''
and Tomsk Polytechnic University Competitiveness Enhancement Program (Russia).
The work of A.~S.~Zh. was supported by the Tomsk State University 
competitiveness improvement program.

\appendix
\section{Contributions of new particles to the muon magnetic 
moment,  proton charge radius  and EDM of fermions}

Contributions of new particles to the anomalous 
magnetic moments of proton and charged leptons read
\begin{widetext}
\eq
 \delta a_S^\psi  =  
   \frac{1}{8 \pi^2} \int\limits_0^1 dx
\frac{(1-x)^2 \, \Big((g_S^\psi)^2-(f_S^\psi)^2 + 
x \Big[(g_S^\psi)^2+(f_S^\psi)^2\Big]\Big)}
{(1-x)^2 + x (\mu_S^\psi)^2}\,,  
\en 
\eq 
\delta a_P^\psi  = 
 - \frac{1}{8 \pi^2} \int\limits_0^1 dx
\frac{(1-x)^2 \, \Big((g_P^\psi)^2-(f_P^\psi)^2 - 
x \Big[(g_P^\psi)^2+(f_P^\psi)^2\Big]\Big)}
{(1-x)^2 + x (\mu_P^\psi)^2}\,,  
\en 
\eq 
 \delta a_V^\psi =  
 \frac{1}{8 \pi^2} \int\limits_0^1 dx
\frac{2 x (1-x) \, 
\Big((g_V^\psi)^2-3 (f_V^\psi)^2 
- x \Big[(g_V^\psi)^2 + (f_V^\psi)^2\Big]\Big)}   
{(1-x)^2 + x (\mu_V^\psi)^2}\,,  
\en 
\eq
\delta a_A^\psi = 
- \frac{1}{8 \pi^2} \int\limits_0^1 dx
\frac{2 x (1-x) \, 
\Big(3 (g_A^\psi)^2 - (f_A^\psi)^2 
+ x \Big[(g_A^\psi)^2 + (f_A^\psi)^2\Big]\Big)}   
{(1-x)^2 + x (\mu_A^\psi)^2}\,,  
\en
\end{widetext}
where $\mu_H^\psi = m_H/M_\psi$, $\psi = p, \ell^-$.

The expression for the $P$-even couplings of scalar, pseudoscalar,
vector and axial particles to the anomalous magnetic moments
of fermions have been obtained before
in Refs.~\cite{Leveille:1977rc,fayet4,McKeen:2009ny,Pospelov:2008zw,Kahn:2016vjr}. 
Note, that
expressions of the $S$, $P$, and $V$ particles are finite,
while the expression for the $A$ is divergent due to longitudinal
part of the axial particle propagator. 
Also divergences due to longitudinal part of the spin-1 particles 
(both vector and axial) occur in the contributions to the proton charge radius. 
As it was shown in Ref.~\cite{Leveille:1977rc} 
(see also Refs.~\cite{fayet4}) 
consideration of the vector and axial particles in the 
renormalized gauge field theory allows to take into account their 
longitudinal part. In particular that implies the cancellation of  divergences 
for the scenario with ultra-violet completion~\cite{Kahn:2016vjr}.  
Here we use phenomenological Lagrangians and restrict 
to use the standard Feynman propagator for spin-0 particles
$D_{J=0}(k^2) = 1/(M^2 - k^2)$ and the one without longitudinal part
for spin-1 particles $D^{\mu\nu}_{J=1}(k^2) = - g^{\mu\nu}/(M^2 - k^2)$. 
 
Below we list the corrections from new particles $(S,P,V,A)$
to the $\langle r_p^2 \rangle^2$:

\begin{widetext}
\eq
\langle \delta r_p^E \rangle^2_S &=&
\frac{1}{8 \pi^2 M_p^2} 
\int\limits_0^1 dx
\frac{(1-x)^2 \, \Big(2 (g_S^p)^2 - (f_S^p)^2
+ x \Big[(g_S^p)^2 + (f_S^p)^2\Big]\Big)}
{(1-x)^2 + x (\mu_S^p)^2}\,, \\
\langle \delta r_p^E \rangle^2_P &=&
- \frac{1}{8 \pi^2 M_p^2} \int\limits_0^1 dx
\frac{(1-x)^2 \, \Big((g_P^p)^2 - 2 (f_P^p)^2
- x \Big[(g_P^p)^2 + (f_P^p)^2\Big]\Big)}
{(1-x)^2 + x (\mu_P^p)^2}\,, \\
\langle \delta r_p^E \rangle^2_V &=&
 \frac{1}{8 \pi^2 M_p^2} \int\limits_0^1 dx
\frac{(1-x) \, \Big((g_V^p)^2 + (f_V^p)^2
+ x \Big[7 (g_V^p)^2 - 6 (f_V^p)^2\Big] 
- 2 x^2 (g_V^p)^2 
- x^3 (f_V^p)^2\Big)}
{(1-x)^2 + x (\mu_V^p)^2}\,, \\
\langle \delta r_p^E \rangle^2_A &=&
- \frac{1}{8 \pi^2 M_p^2} \int\limits_0^1 dx
\frac{(1-x) \, \Big(- (g_V^p)^2 - (f_V^p)^2
+ x \Big[6 (g_A^p)^2 - 7 (f_A^p)^2\Big]
+ 2 x^2 (f_A^p)^2 
+ x^3 (g_A^p)^2\Big)}
{(1-x)^2 + x (\mu_A^p)^2}\,.
\en
\end{widetext}

Lets consider two limiting cases: 
(1) $m_H = m_S = m_P = m_V = m_A \ll M_\psi$, 
(2) $m_H = m_S = m_P = m_V = m_A \gg M_\psi$, 
where $\psi = p, \mu$.  
The total contribution  of new particles into $a^\mu$ and 
proton charge radius read: 

Scenario (1):  
\eq 
\delta a^\mu_{\rm tot} &=& 
\frac{1}{16 \pi^2} \, 
\biggl[ g_a^{(1)} \,-\,  8 \, h_a^{(1)} \, \log(\mu_{H}^{\mu})^2 \biggr]\,, 
\nonumber\\ 
g_a^{(1)} &=& 
3 \Big((g_S^\mu)^2 \,+\, (f_P^\mu)^2\Big) 
  -  \Big((g_P^\mu)^2 \,+\, (f_S^\mu)^2\Big) 
   \nonumber\\
 &+& 2 \Big((g_V^\mu)^2 \,+\,  (f_A^\mu)^2\Big) 
  + 18 \Big((g_A^\mu)^2 \,+\, f_V^\mu)^2\Big)
\,, \nonumber\\
h_a^{(1)} &=&  (f_V^\mu)^2 \,+\, (g_A^\mu)^2 \,,
\en 
\eq 
\langle \delta r_p^E \rangle^2_{\rm tot} &=& 
\frac{1}{16 \pi^2 M_p^2} \ 
\biggl[ g_r^{(1)} \,+\, 6 \, h_r^{(1)} \, \log(\mu_{H}^{p})^2 \biggr]\,, 
\nonumber\\
g_r^{(1)} &=& 
5 \, \Big((g_S^p)^2 \,+\, (f_P^p)^2\Big)  
\,+\, \Big((f_S^p)^2 \,+\, (g_P^p)^2\Big) \nonumber\\
&-&   8 \Big((g_V^p)^2 + (f_A^p)^2\Big)  
\,+\, \frac{47}{3} \Big((f_V^p)^2 + (g_A^p)^2\Big) \,,
\nonumber\\
h_r^{(1)} &=& 
(g_V^p)^2 \,-\, (g_A^p)^2 \,+\, (f_A^p)^2 \,-\, (f_V^p)^2 \,. 
\en

Scenario (2):  
\eq 
\delta a^\mu_{\rm tot} &=& 
\frac{1}{16 \pi^2 (\mu_{H}^{\mu})^2} \, 
\biggl[ g_a^{(2)} \,+\, h_a^{(2)} \, \log(\mu_{H}^{\mu})^2 \biggr]\,, 
\nonumber\\ 
g_a^{(2)} &=&  - \frac{7}{6} \, \Big( 
(g_S^\mu)^2 \,+\, f_P^\mu)^2 \Big)  
\,+\, \frac{11}{6} \, \Big( (g_P^\mu)^2 
\,+\, (f_S^\mu)^2 \Big) \nonumber \\
&+&\frac{2}{3}  \Big( (g_V^\mu)^2 \,+\, (f_A^\mu)^2 \Big) \,-\, 
\frac{10}{3} \Big( (g_A^\mu)^2 \,+\, (f_V^\mu)^2 \Big) 
\,,\nonumber\\
h_a^{(2)} &=& (g_S^\mu)^2 \,-\, (g_P^\mu)^2 
\,-\, (f_S^\mu)^2 \,+\, (f_P^\mu)^2 
\,. 
\en 
\eq 
\langle \delta r_p^E \rangle^2_{\rm tot} &=& 
\frac{1}{8 \pi^2 m_H^2} \ 
\biggl[ g_r^{(2)} \,+\, h_r^{(2)} \, \log(\mu_{H}^{p})^2 \biggr]\,, 
\nonumber\\
g_r^{(2)} &=& 
-\, \frac{8}{3} \, \Big((g_S^p)^2 + (f_P^p)^2 \Big)\nonumber \\
&+& \frac{11}{6} \, \Big((g_P^p)^2 + (f_S^p)^2 \Big)   
+ \frac{13}{6}  \Big((g_V^p)^2 + (f_A^p)^2 \Big)  \nonumber \\
&-& \frac{49}{2} \, \Big((g_A^p)^2 \,+\, (f_V^p)^2 \Big)   
\,, \nonumber\\
h_r^{(2} &=& 
2 \, \Big( (g_S^p)^2 \,+\, (f_P^p)^2\Big) - 
\Big( (g_P^p)^2 + (f_S^p)^2\Big)\nonumber\\
&+&\Big( (g_V^p)^2 +(f_V^p)^2 \,+\, 
      (g_A^p)^2 + (f_A^p)^2\Big). 
\en

The contributions of individual diagrams in Fig.~\ref{fig:diag2} 
are given by: 

Diagrams 1+2: 

$S(P)$-boson exchange 
\eq 
d^E_{I} = \frac{e g_I f_I}{8\pi^2 M_\psi}  \, I(\mu_I^\psi) \,, 
\quad\quad I = S, P \,. 
\en 

Diagrams 3+4: 

$SV$-boson exchange 
\eq 
d^E_{SV} = \frac{e g_{\gamma SV} g_V f_S}{16\pi^2 M_p} \, 
J(\mu_S^\psi,\mu_V^\psi) \,.  
\en 

$PA$-boson exchange 
\eq 
d^E_{PA} = \frac{e g_{\gamma PA} g_P f_A}{16\pi^2 M_p} \, 
J(\mu_P^\psi,\mu_A^\psi) \,.  
\en 

Diagrams 5+6: 

$PV$-boson exchange 
\eq 
d^E_{PV} = \frac{e f_{\gamma PV} g_P g_V}{16\pi^2 M_p} \, 
J(\mu_P^\psi,\mu_V^\psi) \,.
\en 
Here we introduced the structure integrals 
\eq
I(\mu) = \int\limits_0^1 dx \, \frac{x^2}{x^2 + (1-x) \mu^2} 
\en
for diagrams 1,2 and 
\eq 
J(\mu,\tau) = \frac{1}{\mu^2-\tau^2} \, \int\limits_0^1 dx \, x^2 \, 
\log\frac{x^2 + \mu^2 (1-x)}{x^2 + \tau^2 (1-x)}
\label{Jmutau}
\en 
for diagrams 3-6.  
For equal masses of bosons, i.e. for $\mu=\tau$ the loop integral 
$J(\mu,\tau)$ is simplified to 
$J(\mu) = \int_0^1 dx x^2 (1-x)/(x^2 + (1-x)\mu^2)$. 
As before we consider the limits: (1) small fermion masses 
$\mu, \tau  \gg 1$ 
and (2) small boson masses $\mu, \tau  \ll 1$. 
In first case the structure integrals read: 
\eq 
I(\mu) = \log(\mu^2)/\mu^2 
\en 
\eq 
J(\mu,\tau) = \frac{1}{3 (\mu^2 - \tau^2)} \, \log\frac{\mu^2}{\tau^2} \,, 
\quad J(\mu) =  \frac{1}{3 \mu^2} \,, 
\en 
For the second case we get: 
\eq 
I(\mu) = 1\,, \quad J(\mu) = J(\mu,\tau) = \frac{1}{2} \,. 
\en

\end{document}